\documentclass{article}

\def\xxinput#1{\input#1}

\xxinput{vsolj02.sty}

\usepackage{longtable}
\usepackage[dvipdfmx]{graphicx}
\usepackage[comma,colon]{natbib}

\usepackage[OT2,T1]{fontenc}

\def\cite{\citealt}
\setcitestyle{aysep={}}

\newcounter{author}
\def\altaffilmark#1{$^{#1}$}
\def\altaffiltext#1{$^{#1}$\,}

\def\authorcount#1#2{{\refstepcounter{author}\label{#1}
                     \altaffiltext{\ref{#1}}{#2}}}

\begin{document}

\begin{center}

\title{Negative superhumps in the eclipsing Z Cam + VY Scl star ES Dra}

\author{
        Taichi~Kato\altaffilmark{\ref{affil:Kyoto}}
}
\email{tkato@kusastro.kyoto-u.ac.jp}

\authorcount{affil:Kyoto}{
     Department of Astronomy, Kyoto University, Sakyo-ku,
     Kyoto 606-8502, Japan}

\end{center}

\begin{abstract}
\xxinput{abst.inc}
\end{abstract}

\section{Introduction}

   ES Dra was originally selected as a ultraviolet-excess object
(PG 1524$+$622) and was confirmed to be a cataclysmic variable (CV)
by spectroscopy \citep{PGsurvey}.  The orbital period ($P_{\rm orb}$)
and the nature of this object has much been disputed.
\citet{and91esdra} reported weak detections of the periods
1.07(8), 0.53(3) and 0.125(3)~d.  \citet{and91esdra} suggested
a characteristic time-scale of 0.10--0.13~d.
Although \citet{rin93thesis} suggested a spectroscopic orbital
period of 0.179~d, it was finally published in \citet{rin12esdra}.
\citet{mis95PGCV} could not detect variations at the reported
$P_{\rm orb}$ from observations on two nights.  
In the meantime, Tonny Vanmunster detected periods of 0.064(8)
and 0.121(6)~d during the 2001 June outburst.\footnote{
  $<$https://www.cbabelgium.com/cv\_2001/Dra\_ES\_jun\_2001.html$>$.
}  Combined with observations by Jerry Foote, Vanmunster
claimed to detect superhumps with a period of 0.1267(20)~d
and ES~Dra was once considered as an SU UMa-type dwarf nova
above the period gap.  \citet{bak01esdra} reported $\sim$20-d
variations with an amplitude of 1.6~mag.  \citet{bak01esdra}
suggested that the object is a novalike variable.

   \citet{rin12esdra} published a radial-velocity study
giving $P_{\rm orb}$=0.17660(6)~d and a spectral type of
M2$\pm$1 for the secondary.  \citet{rin12esdra} used
American Association of Variable Star
Observers (AAVSO) observations and concluded it to be
a Z Cam star after the identification of standstills in
1995 and 2009.  \citet{sim14zcamcamp} reached the same
conclusion by showing a standstill in 2012 January to August.

\section{Long-term variation}

\begin{figure*}
\begin{center}
\includegraphics[width=16cm]{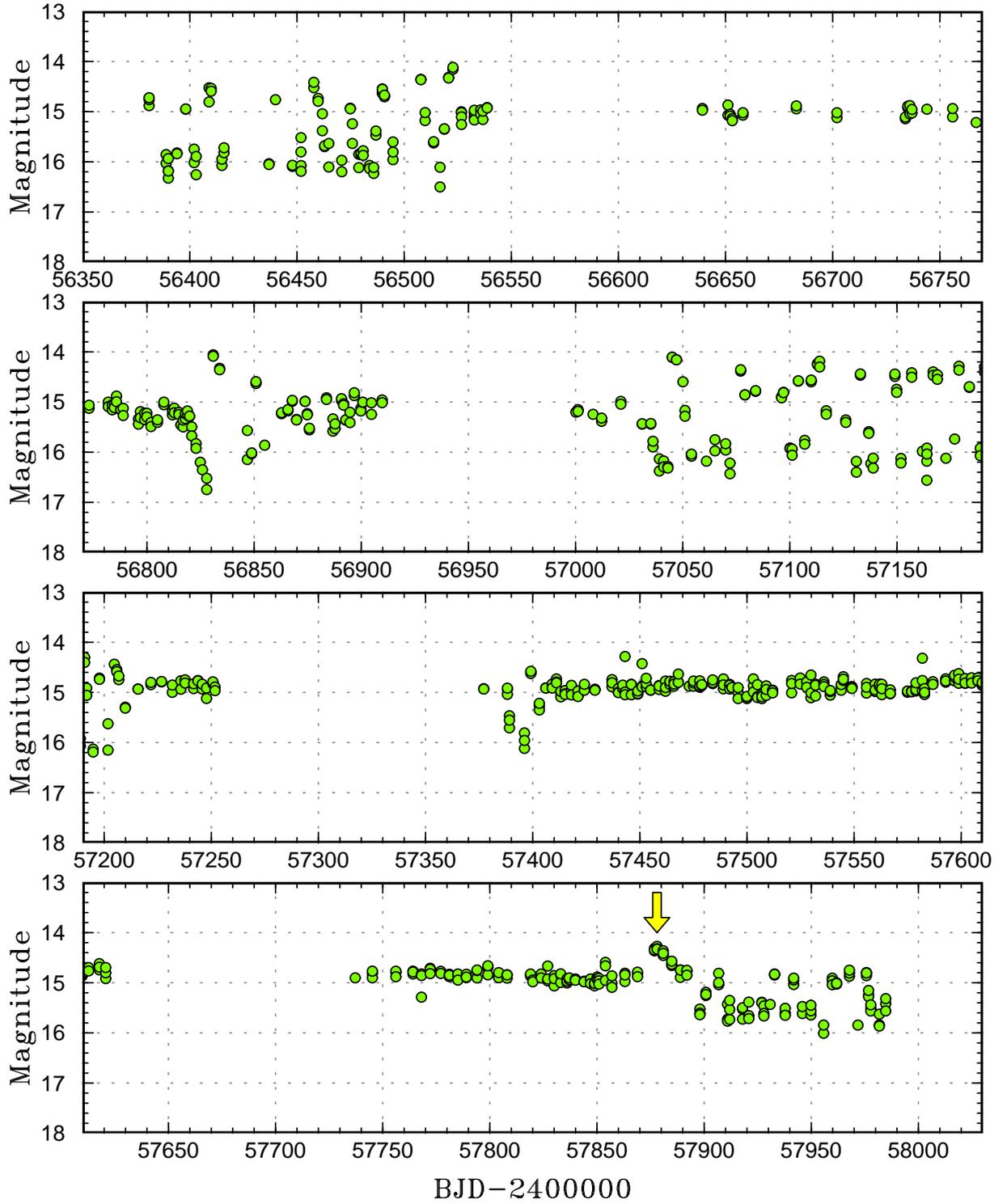}
\caption{
  Long-term light curve of ES Dra using ASAS-SN $V$-band data (1).
  Both states with dwarf nova outbursts and standstills were
  recorded.  The long standstill starting on BJD 2457406
  was terminated by brightening (arrow in the fourth panel).
}
\label{fig:esdralc}
\end{center}
\end{figure*}

\begin{figure*}
\begin{center}
\includegraphics[width=16cm]{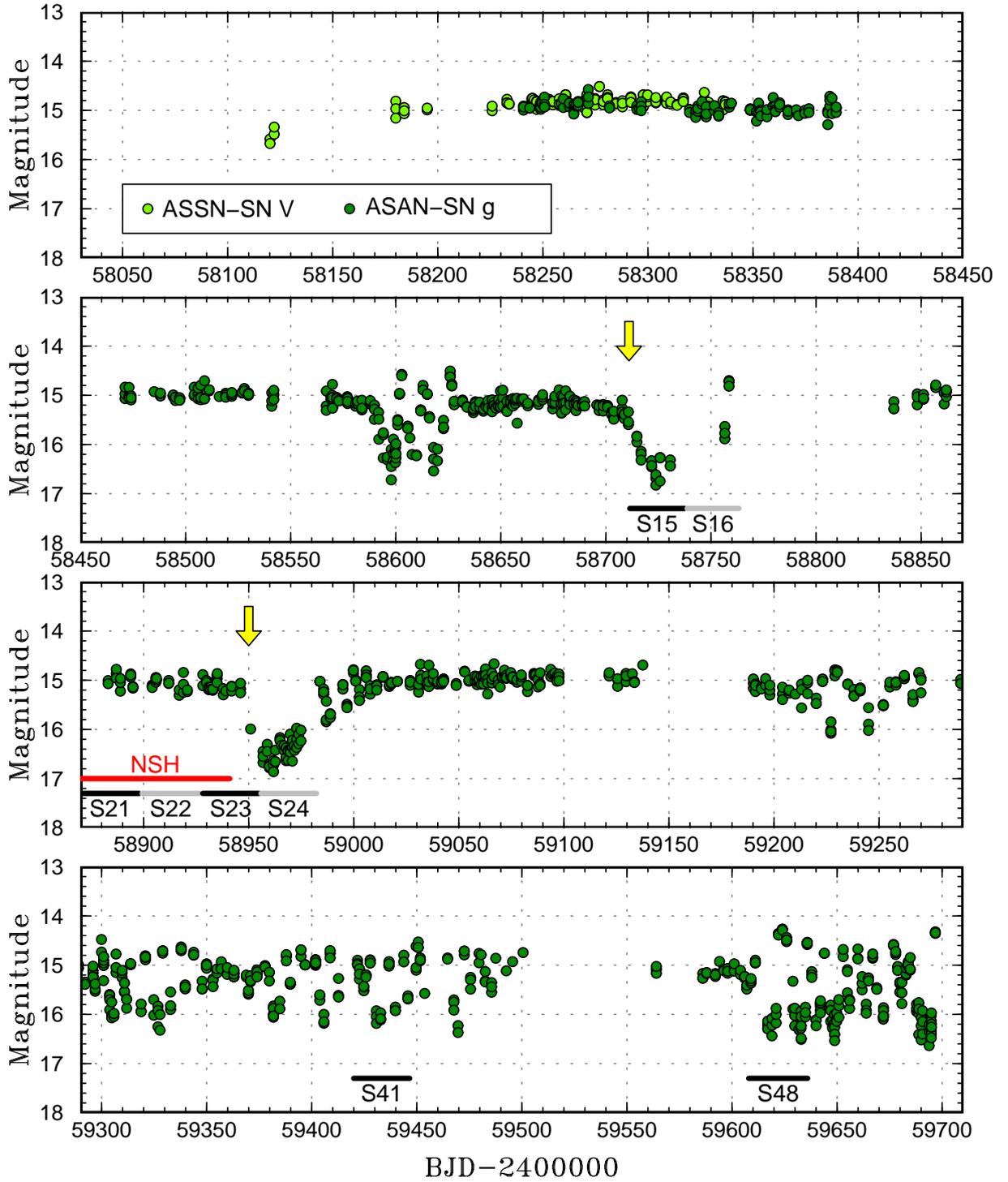}
\caption{
  Long-term light curve of ES Dra using ASAS-SN data (2).
  Two fading episodes following standstills are shown
  by the arrows.
  The horizontal bars represent TESS sectors.
  The red horizontal bar represents the interval when
  negative superhumps (NSH) were present in the TESS data.
}
\label{fig:esdralc2}
\end{center}
\end{figure*}

   I used All-Sky Automated Survey for Supernovae
(ASAS-SN) Sky Patrol data \citep{ASASSN,koc17ASASSNLC}
and I confirmed that ES Dra is indeed a Z Cam star
(vsnet-chat 8114).\footnote{
  $<$http://ooruri.kusastro.kyoto-u.ac.jp/mailarchive/vsnet-chat/8114$>$.
}  Although the 2017 standstill was terminated by
brightening (outburst), not by fading as in ordinary Z Cam stars
\citep[see e.g.][]{szk84AAVSO},
the importance of this finding was not recognized at that time.
In 2020, I noticed two fading episodes
(2019 August and 2020 April) in the ASAS-SN data and concluded
that this object is a Z Cam + VY Scl star (vsnet-chat 8473).\footnote{
  $<$http://ooruri.kusastro.kyoto-u.ac.jp/mailarchive/vsnet-chat/8473$>$.
}  This classification has been adopted in AAVSO
Variable Star Index (VSX: \cite{wat06VSX}).
The long-term behavior based on the ASAS-SN data is shown in
figures \ref{fig:esdralc} and \ref{fig:esdralc2}.

\section{Orbital period and profile}

\begin{figure*}
\begin{center}
\includegraphics[width=16cm]{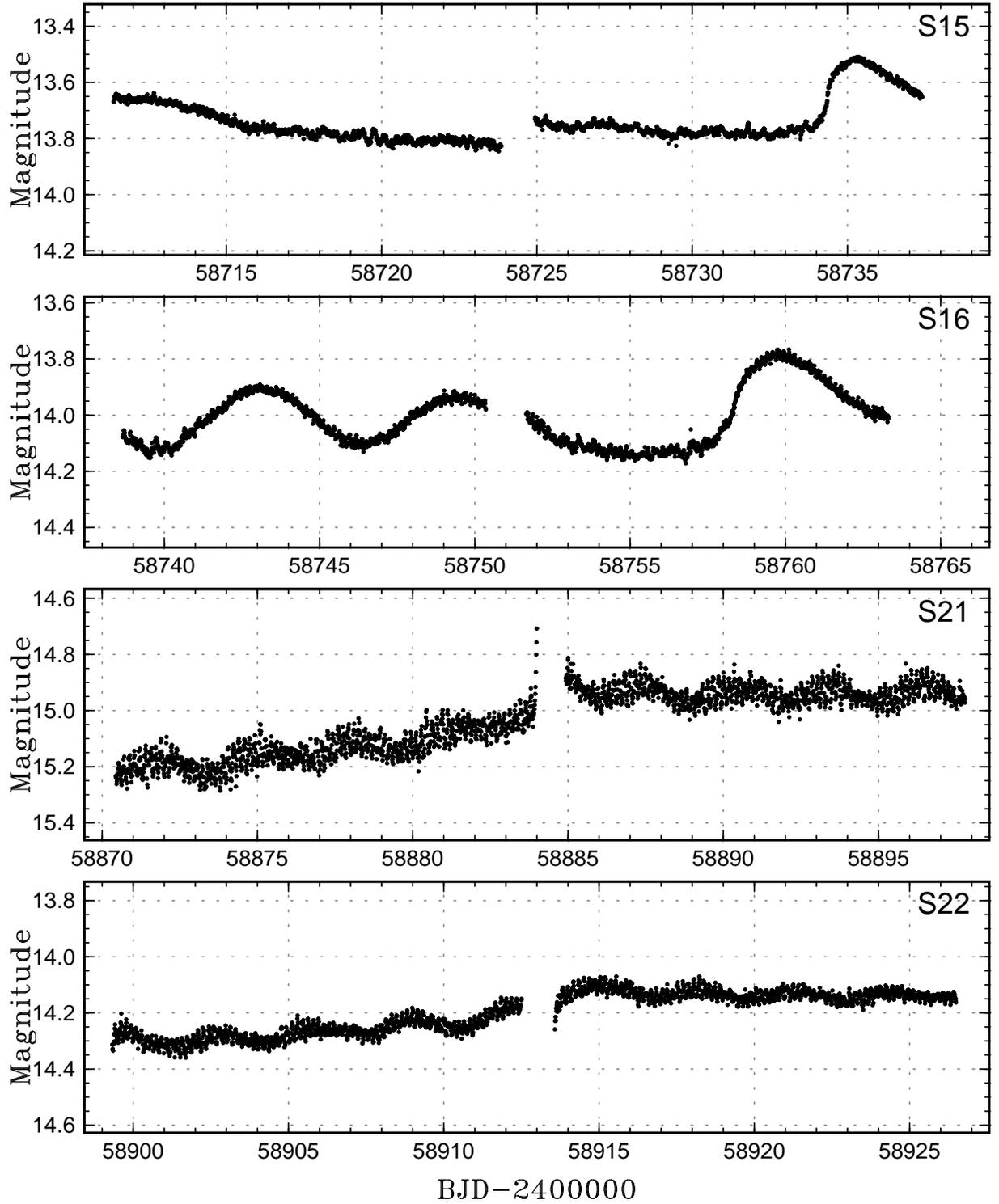}
\caption{
  TESS light curve of ES Dra (1).  The data were binned to 0.01~d.
  The numbers following S represent the TESS sector numbers.
  The magnitudes were defined as $-2.5\log_{10}({\rm flux}/15000)+10$.
  Beat phenomenon with a period of 3~d was present in S21 and S22.
}
\label{fig:esdratess}
\end{center}
\end{figure*}

\begin{figure*}
\begin{center}
\includegraphics[width=16cm]{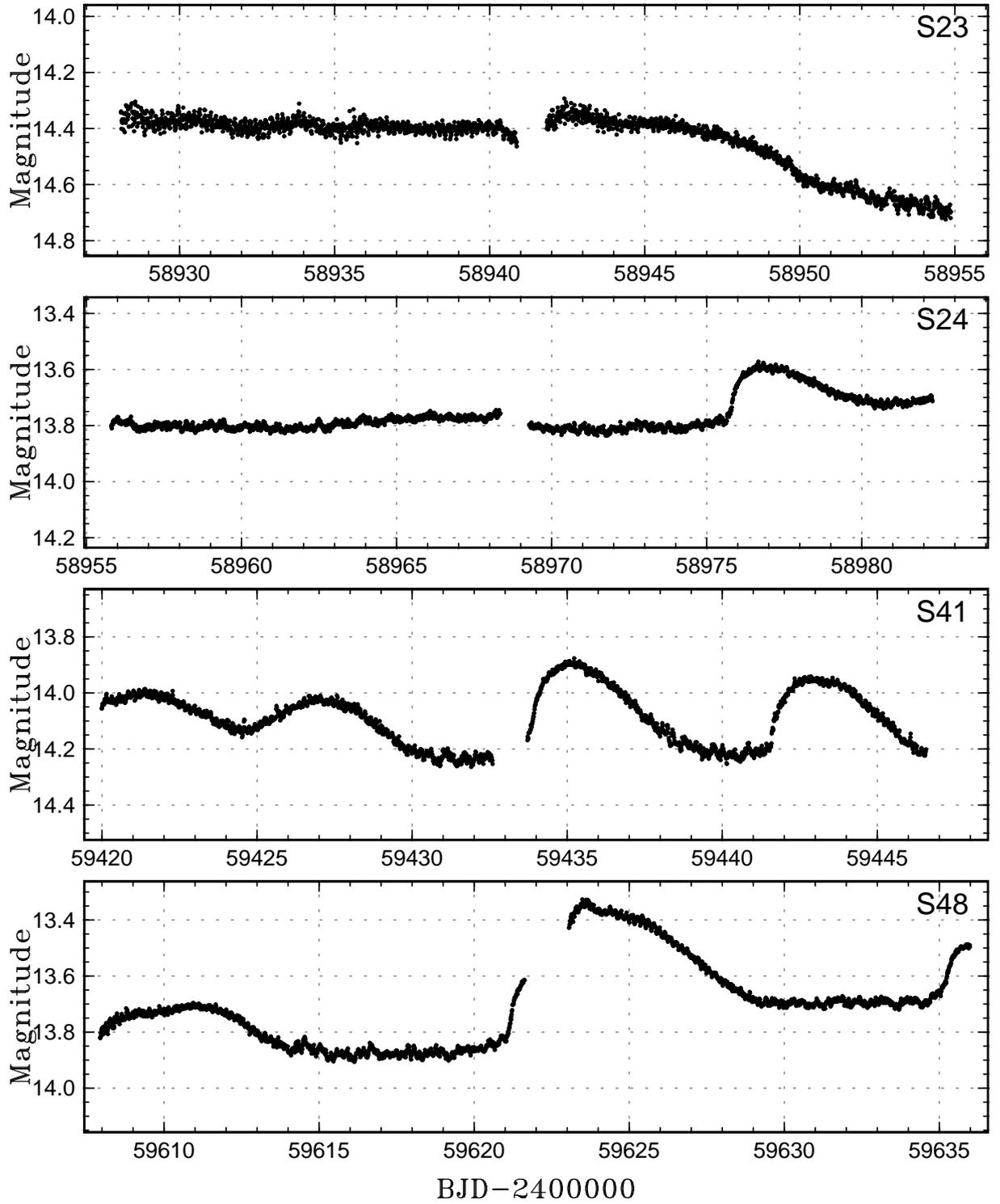}
\caption{
  TESS light curve of ES Dra (2).  See figure \ref{fig:esdratess}
  for explanation.
}
\label{fig:esdratess2}
\end{center}
\end{figure*}

   I also analyzed Transiting Exoplanet Survey Satellite (TESS)
observations \citep{ric15TESS}.\footnote{
  $<$https://tess.mit.edu/observations/$>$.
}  The full light-curve
is available at the Mikulski Archive for Space Telescope
(MAST\footnote{
  $<$http://archive.stsci.edu/$>$.
}).  I used six high level science products (HLSP)
sectors between 2019 August 15 and 2022 February 25
(figures \ref{fig:esdratess}, \ref{fig:esdratess2}).
I obtained $P_{\rm orb}$ for different states
using the Phase Dispersion Minimization (PDM, \cite{PDM})
method after removing long-term trends by locally-weighted
polynomial regression (LOWESS: \cite{LOWESS}).
The errors of periods by the PDM method were
estimated by the methods of \citet{fer89error} and \citet{Pdot2}.
The results are shown in table \ref{tab:porb}.
A PDM analysis of the entire segments as a whole did not
yield a unique period.  This was due to the variation
of the orbital profile as discussed later.
Shallow eclipses were seen following the orbital hump
in S16, S21--24 and S48 and I obtained a unique period of
0.1774992(2)~d after combination of these segments.
The orbital period was refined using
the Markov-Chain Monte Carlo (MCMC)-based
method introduced in \citet{Pdot2}.
The resultant ephemeris is
\begin{equation}
{\rm Min (BJD)} = 2459148.9510(7) + 0.17749895(17) E.
\label{equ:esdraecl}
\end{equation}

\begin{table}
\caption{Orbital period of ES Dra from TESS data}\label{tab:porb}
\begin{center}
\begin{tabular}{ccccc}
\hline
Interval BJD$-$2400000 & TESS sector & Period (d) & Amplitude (mag) \\
\hline
58711--58737 & S15 & 0.17746(4) & 0.006 \\
58738--58763 & S16 & 0.17750(2) & 0.015 \\
58870--58982 & S21--24 & 0.177491(8) & 0.013 \\
59419--59446 & S41 & 0.17752(1) & 0.014 \\
59607--59636 & S48 & 0.17749(2) & 0.008 \\
\hline
\end{tabular}
\end{center}
\end{table}

\begin{figure*}
\begin{center}
\includegraphics[width=16cm]{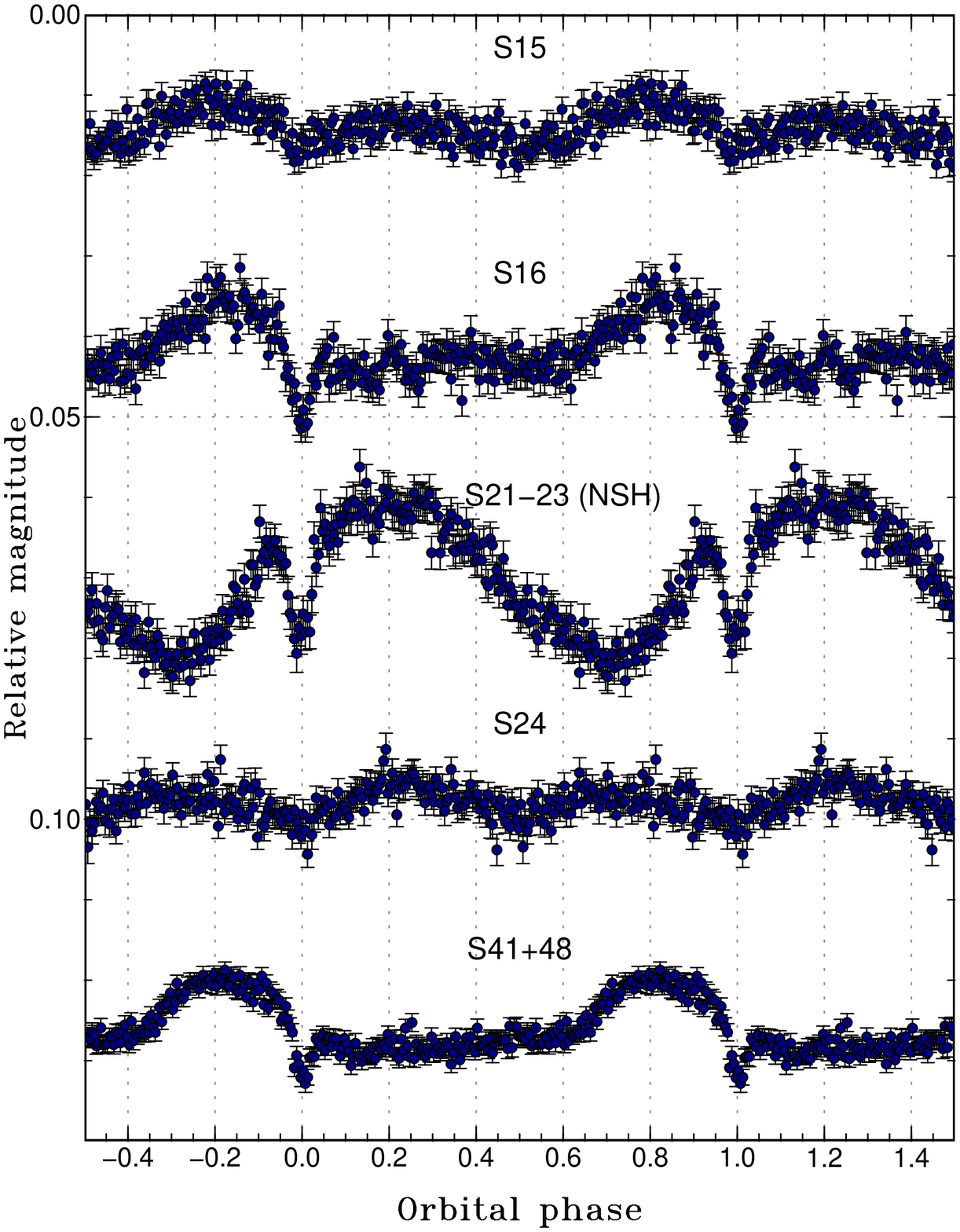}
\caption{
  Mean orbital profiles of ES Dra.  The ephemeris in
  equation (\ref{equ:esdraecl}) was used.
  The numbers following S represent TESS sector numbers
  (see figure \ref{fig:esdralc2}).
  S21--23 (NSH) represents the interval when negative superhumps
  were detected.
  S24 includes a short segment of S23 when negative superhumps
  were absent.
}
\label{fig:prof}
\end{center}
\end{figure*}

   As will be shown later, this object showed negative superhumps
between 2020 January and March and the orbital profile was
variable depending on the presence/absence of negative superhumps.
The mean orbital profiles in different states
are shown in figure \ref{fig:prof}.
When negative superhumps were completely absent, the orbital
profile was that of a dwarf nova with a shallow eclipse
showing an orbital hump at phase 0.8 (S16, S41 and S48).
When negative superhumps were prominent, the hump
moved to a different phase (S21--23 NSH).
This result would, however, give a somewhat artificial impression
since the amplitudes of negative superhumps were larger than
those of orbital humps and the hump phase defined by $P_{\rm orb}$
may not be very adequate.
This curve, however, indicates that the eclipses were present
at the correct orbital phase even during the phase of
negative superhumps and confirms the validity of $P_{\rm orb}$
determined in this study.  This variation of the profile
was the reason why a PDM anlysis of the entire data set did not
yield a unique $P_{\rm orb}$.  The S15 and S24 curves in this
figure correspond to the epochs when the object experienced
fading episodes (figure \ref{fig:esdralc2}).
It was most likely that the disk shrunk and eclipses disappeared.
The orbital profile resembled that of ellipsoidal variation,
consistent with a VY Scl star in low state.

\section{Negative superhumps}

\begin{figure*}
\begin{center}
\includegraphics[width=16cm]{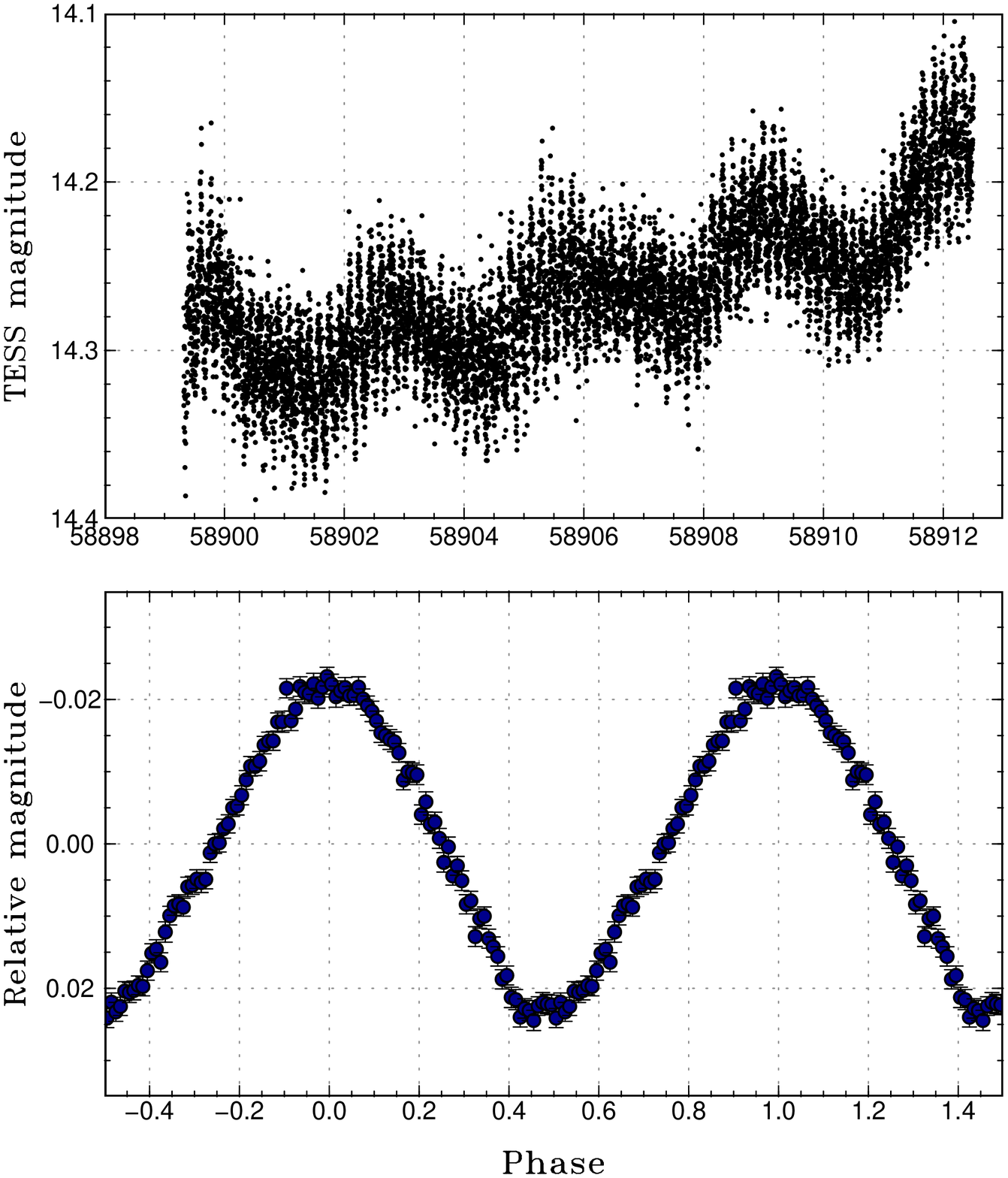}
\caption{
  Negative superhumps in ES Dra.
  (Upper:) Example of TESS light curve.
  Both short-period modulations
  (negative superhumps) and the 3-d beat period were present.
  (Lower:) Profile of negative superhumps.  The zero phase
  and the period were defined as BJD 2458905.118 and 0.167830~d,
  respectively.
}
\label{fig:nsh}
\end{center}
\end{figure*}

\begin{figure*}
\begin{center}
\includegraphics[width=16cm]{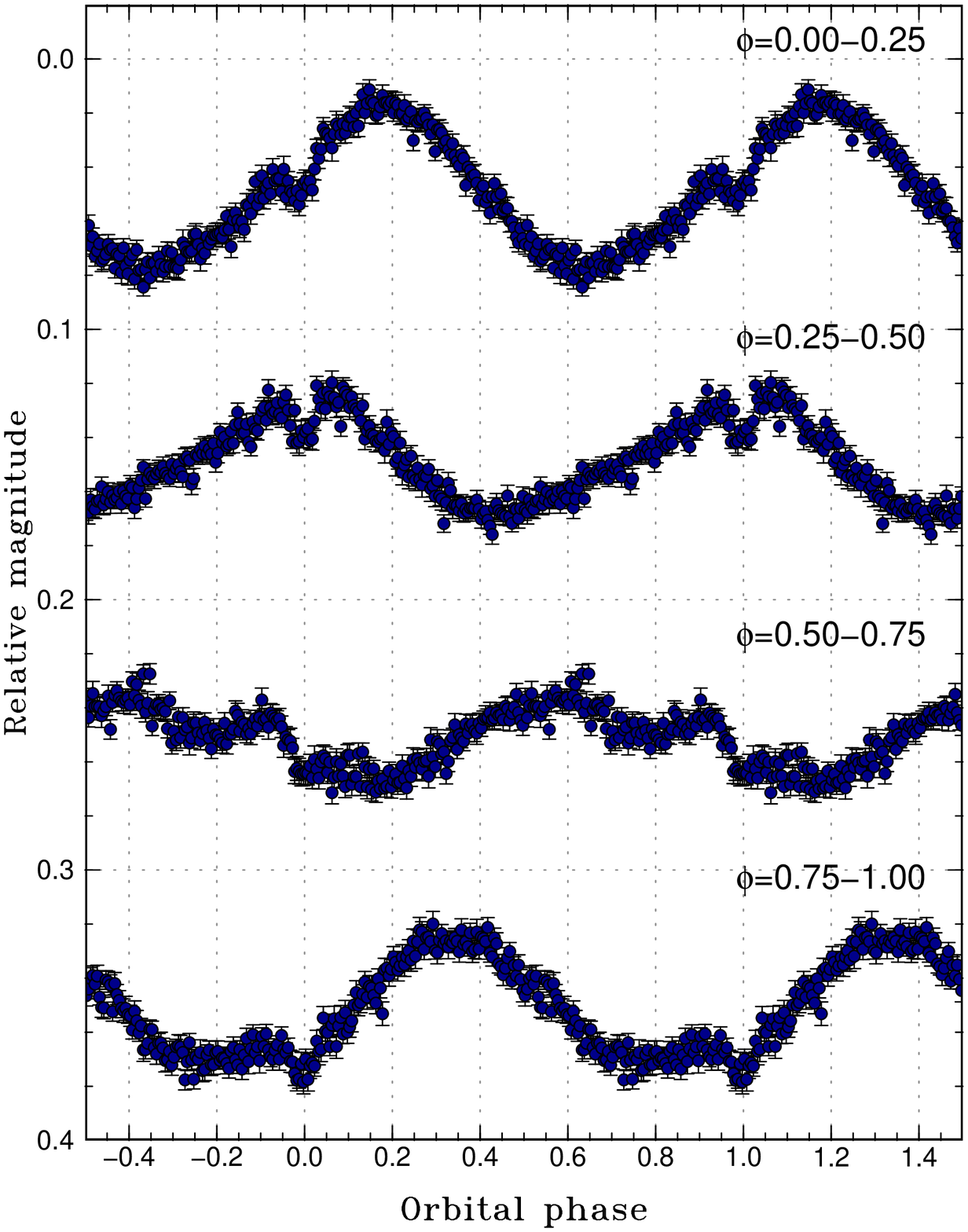}
\caption{
  Variation of orbital profiles of ES Dra during
  the phase of negative superhumps.
  The beat phases $\phi$ were determined
  using the epoch of BJD 2458905.118 (see figure \ref{fig:nsh})
  and the beat period of 3.08096~d.
  The orbital phases were defined by equation (\ref{equ:esdraecl}).
}
\label{fig:beat}
\end{center}
\end{figure*}

   A PDM analysis of the segment BJD 2458870--2458940
yielded a period of negative superhumps of 0.167830(2)~d.
The beat period with $P_{\rm orb}$ (3.08~d) is clearly
visible in the TESS light curve (figure \ref{fig:nsh}).
The variation of orbital profiles depending on
the beat phase (i.e. orientation of the disk to the observer
assuming a tilted disk) is shown in figure \ref{fig:beat}.
Although eclipses were present in most beat phases,
they were not apparent for $\phi$=0.50--0.75.

   The fractional superhump deficit
for negative superhumps is $\epsilon^-$ = $-$5.4\%,
which is a typical value for this $P_{\rm orb}$
\citep{woo09negativeSH}.  It is widely believed that negative
superhumps arise from the variable release of the potential
energy when accreting on a precessing, tilted disk
\citep{woo00SH,mur02warpeddisk,woo07negSH}.  Recent detailed
analysis of Kepler observations provided a strong support
to this interpretation \citep{kim20kic9406652,kim21kic9406652}.
ES Dra showed a clear beat phenomenon and this is also
a support to the precessing, tilted disk as the origin
of negative superhumps.  A similar case has also been
reported using TESS observations of LS Cam
\citep{ste21lscam,raw22lscamv902monj0746}.
Eclipses in ES Dra are grazing and the secondary only eclipses
the outermost part of the disk.
The absence of eclipses in for $\phi$=0.50--0.75 suggests
that the part of the disk facing the secondary
was most distant from the secondary when the secondary
passes in front of the disk in this beat phase
and was most difficult to eclipse.
This variation of the eclipse depth depending on
the beat phase also supports the precessing, tilted disk
as the interpretation of negative superhumps.

\begin{figure*}
\begin{center}
\includegraphics[width=16cm]{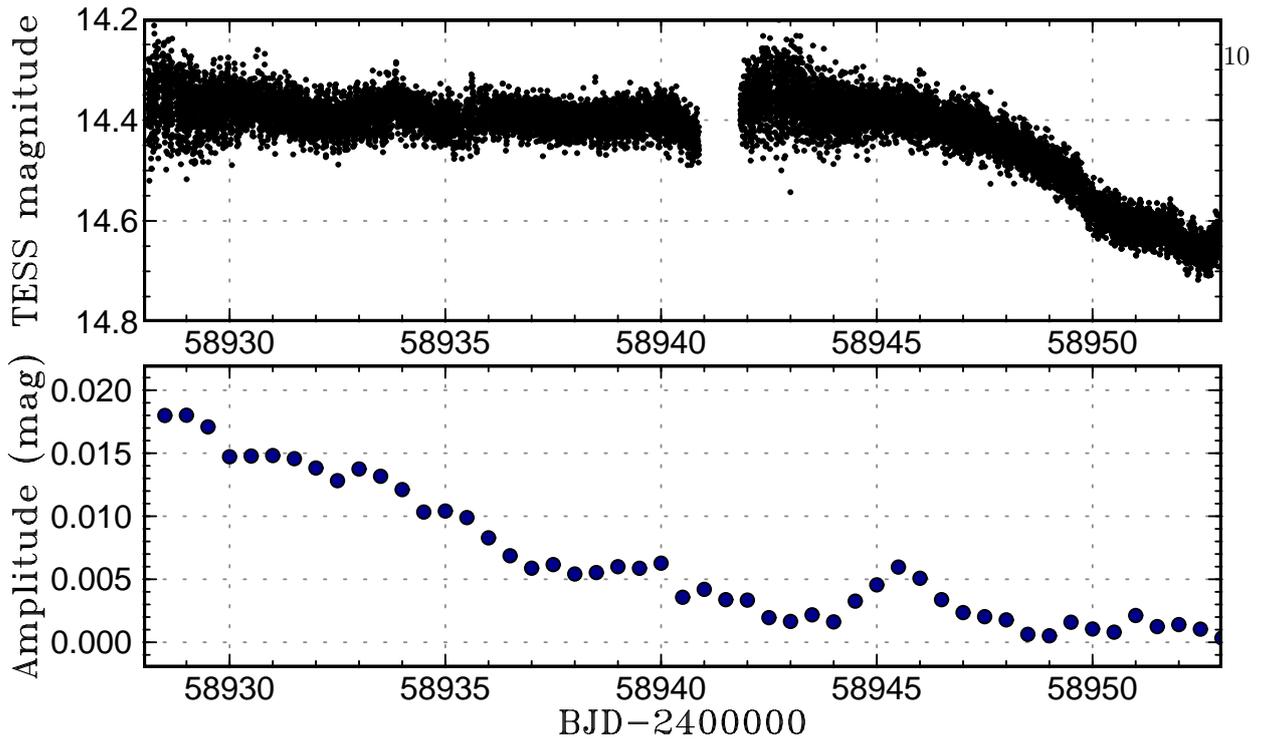}
\caption{
  Light variation and amplitude of negative superhumps
  when ES Dra entered a VY Scl-type low state.
  (Upper:) TESS light curve.
  (Lower:) Amplitude of negative superhumps.  The values were
  obtained for 3-d segments considering the beat phenomenon.
  The amplitude reached almost zero on BJD 2458943.
  The fading started on BJD 2458947.
}
\label{fig:s23}
\end{center}
\end{figure*}

   The disappearance of negative superhumps when the object
entered the low state was probably caused by the decrease
of the mass-transfer rate.  Negative superhumps disappeared
4~d earlier than the fading episode started
(see figure \ref{fig:s23}).  It was possible that the
mass-transfer rate quickly dropped and the spot on
the disk disappeared 4~d before
the total luminosity of the disk started to decline.

\section{System parameters and absolute magnitude}

   Assuming a standard Roche-filling secondary, the mass
of the secondary ($M_2$) is expected to be 0.35$M_\odot$
\citep{kni06CVsecondary}.  The spectral type (M3.3)
is in agreement with M2$\pm$1 by \citet{rin12esdra}.
The white dwarf in ES Dra is not expected to be
particularly massive as judged from the absence of
high excitation lines \citep{rin12esdra}.
Assuming an average-mass white dwarf ($M_1$=0.81$M_\odot$)
for CVs \citep{pal22CVparam}, the mass ratio is expected
to be $q=M_2/M_1=0.43$.  The $K_1$ velocity of the primary
is expected to be 112 km s$^{-1}$.
Assuming that the disk of ES Dra in
high state (in standstill) has a radius reaching
the tidal truncation radius
\begin{equation}
\label{equ:rtidal}
r_{\rm tidal} = \frac{0.60}{1+q},
\end{equation}
in unit of the binary separation ($A$) \citep{pac77ADmodel},
and assuming an optically thick standard disk, the observed eclipse
could be best modelled by an inclination of
$i$=62.0$\pm$0.5$^\circ$.  This value is most dependent on
the disk radius and $i$=64$^\circ$ is the best for
a disk radius of 0.38$A$.

   The $K$ velocity of the emission
($K_{\rm em}$=134$\pm$11 km s$^{-1}$)
in \citet{rin12esdra} is slightly larger than the value
expected for these system parameters.  If the emission
line exactly traces the motion of the primary
($K_{\rm em} = K_1 \sin i$), the best parameters are
$M_1$=0.53$M_\odot$ and $i$=61$^\circ$.  It is well-known
that such an ideal case is rarely achieved, and we would better
rely on the first solution until $M_1$ or $q$ is independently
measured.

   Using Gaia EDR3 \citep{GaiaEDR3}, the absolute magnitude
$M_V$ at standstill (in average) is $+$5.7 and
peaks of dwarf nova outbursts reach $+$4.9.
For an object with $i$=61--64$^\circ$, the effect of
inclination
\begin{equation}
\label{equ:incldepend}
\Delta M_v(i) = -2.5 \log_{10} \left[ (1 + \frac{3}{2}\cos i) \cos i \right],
\end{equation}
is 0.19--0.35~mag.  By adopting $i$=62.0$^\circ$, the corrected
$M_V$ for standstill and outburst peak are $+$5.5 and $+$4.7,
respectively.

\section{Standstill terminated by brightening}

   The phenomenon of a standstill terminated by brightening
(outburst) in 2017 May is a signature of an IW And star
\citep{sim11zcamcamp1,ham14zcam,kat19iwandtype}.
As described in \citet{kat19iwandtype}, typical IW And stars
show a tendency of recurrence of this phenomenon which is
frequently followed by a dip and subsequent damping oscillations.
Such a cycle has not been yet apparent in ES Dra
and it would be interesting to see whether this object
shows cyclic behavior in future as in typical IW And stars
or the 2017 May event was a sporadic one and its resemblance
to the IW And-type phenomenon was superficial.

\section*{Acknowledgements}

This work was supported by JSPS KAKENHI Grant Number 21K03616.
The author is grateful to the ASAS-SN and TESS teams
for making their data available to the public.
I am grateful to Naoto Kojiguchi for helping downloading
the TESS data.

\section*{List of objects in this paper}
\xxinput{objlist.inc}

\section*{References}

We provide two forms of the references section (for ADS
and as published) so that the references can be easily
incorporated into ADS.

\renewcommand\refname{\textbf{References (for ADS)}}

\newcommand{\noop}[1]{}\newcommand{\hyphalt}{-}

\xxinput{esdraaph.bbl}

\renewcommand\refname{\textbf{References (as published)}}
\xxinput{esdra.bbl.vsolj}

\end{document}